\documentclass[10pt,conference]{IEEEtran}
\IEEEoverridecommandlockouts
\usepackage{epsfig,graphicx,subfigure,psfrag,amsmath,cases,bm}
\usepackage{latexsym,amssymb,algorithm,mathtools}
\usepackage{algorithmic}
\usepackage{color}
\usepackage{url}
\usepackage{scrtime}
\usepackage{stfloats}
\usepackage{tablefootnote}
\usepackage{cite}
\usepackage{amsfonts,mathabx}
\usepackage{textcomp}
\usepackage{xcolor,capt-of}
\usepackage{multirow}
\usepackage{amsthm}
\usepackage{geometry}
\def\BibTeX{{\rm B\kern-.05em{\sc i\kern-.025em b}\kern-.08em
    T\kern-.1667em\lower.7ex\hbox{E}\kern-.125emX}}
\makeatletter
\newcommand*{\rom}[1]{\expandafter\@slowromancap\romannumeral #1@}
\makeatother

\DeclareMathOperator{\mino}{minimize}

\newtheorem{lemma}{Lemma}
\IEEEaftertitletext{\vspace{-2em}}
\allowdisplaybreaks

\title{Globally Optimal Resource Allocation Design for IRS-Assisted Multiuser Networks with Discrete Phase Shifts\vspace{-0.7em}}
\author{\IEEEauthorblockN {Yifei Wu\IEEEauthorrefmark{1}, Dongfang Xu\IEEEauthorrefmark{2}, Derrick Wing Kwan Ng\IEEEauthorrefmark{3}, Robert Schober\IEEEauthorrefmark{1}, and Wolfgang Gerstacker\IEEEauthorrefmark{1}}

\IEEEauthorrefmark {1}Friedrich-Alexander-Universit\"at
Erlangen-N\"urnberg, Germany\\
\IEEEauthorrefmark {2}Dept. of ECE, The Hong Kong University of Science and Technology, Hong Kong\\
\IEEEauthorrefmark {3}The University
of New South Wales, Australia

}

\begin{document}
\maketitle
\begin{abstract}
    Intelligent reflecting surfaces (IRSs) are envisioned as a low-cost solution to achieve high spectral and energy efficiency in future communication systems due to their ability to customize wireless propagation environments. Although resource allocation design for IRS-assisted multiuser wireless communication systems has been exhaustively investigated in the literature, the optimal design and performance of such systems are still not well understood. To fill this gap, in this paper, we study optimal resource allocation for IRS-assisted multiuser multiple-input single-output (MISO) systems. In particular, we jointly optimize the beamforming at the base station (BS) and the discrete IRS phase shifts to minimize the total transmit power. For attaining the globally optimal solution of the formulated non-convex combinatorial optimization problem, we develop a resource allocation algorithm with guaranteed convergence based on Schur's complement and the generalized Bender's decomposition. Our numerical results reveal that the proposed algorithm can significantly reduce the BS transmit power compared to the state-of-the-art suboptimal alternating optimization-based approach, especially for moderate-to-large numbers of IRS elements.
\end{abstract}
\vspace*{-2mm}
\section{Introduction}
\vspace*{-2mm}
Recently, intelligent reflecting surfaces (IRSs) have attracted considerable research interest from both academia and industry due to their unique and attractive characteristics. In particular, given their adaptive reflecting elements, IRSs can be flexibly programmed to control the reflections of the propagating wireless signals. Moreover, commonly deployed as thin rectangular planes, IRSs can be attached to building facades, indoor ceilings, and vehicles, hence introducing extra design degrees of freedom (DoFs) for improving the performance of wireless communication systems, e.g., expanding the service coverage, enhancing the security of communications, etc. As a result, IRSs have been envisioned as a promising enabler for establishing future ubiquitous high-data-rate communication networks. Motivated by the aforementioned appealing properties, numerous works have studied the integration of IRSs with other advanced communication techniques, e.g., multiple-input multiple-output (MIMO) transmission, simultaneous wireless information and power transfer (SWIPT) \cite{xu2022optimal}, and physical layer security \cite{9014322}.

To unleash the potential of IRS-aided wireless communications, several works have focused on the joint design of the transmit beamforming at the base station (BS) and the IRS phase shifters. For instance, in \cite{9154337}, an iterative algorithm based on the branch-and-bound (BnB) method was developed to obtain the globally optimal solution for a single-user IRS-assisted multiple-input single-output (MISO) system. Also, the authors in \cite{9014322} adopted the majorization-minimization (MM) and block coordinate descent (BCD) techniques for the design of a single-user IRS-assisted secure wireless communication system. However, \cite{9014322} and \cite{9154337} made the overly optimistic assumption that the phases of the IRS phase shifters are continuous, which is impractical due to the resulting high implementation cost, in terms of energy consumption and hardware, and the delicate design of high-resolution reflecting elements. In fact, practical large-scale IRSs usually employ 1-bit on-off phase shifters or 2-bit quadruple-level phase shifters, as these architectures are more cost-effective \cite{wu2019beamforming}. Some initial works have considered discrete IRS phase shifters, e.g., \cite{wu2019beamforming} and \cite{shi2022multiuser}. For instance, the authors in \cite{wu2019beamforming} focused on a single-user scenario and determined the optimal BS beamformer and discrete IRS phase shifts by employing an enumeration-based algorithm. Yet, this result is not applicable to multiuser systems since the optimal multiuser beamforming matrix cannot be obtained in closed form. Besides, in \cite{shi2022multiuser}, a suboptimal algorithm based on alternating optimization (AO) was developed to promote user fairness for an IRS-assisted multiuser system with discrete phase shifts. However, the AO-based algorithm in \cite{shi2022multiuser} cannot guarantee joint optimality of the BS beamformer and the IRS phase shifts, as its performance highly depends on the selection of the initial point, which may result in unsatisfactory performance \cite{9014322}, \cite{xu2021resource}. Thus, in this paper, we investigate for the first time the jointly optimal BS beamforming and IRS reflection coefficient design for multiuser IRS-assisted systems with discrete phase shifters to fully reveal the potential of IRS-assisted systems. The main contributions of this paper can be summarized as follows:
\vspace*{-1.5mm}
\begin{itemize}
    \item To handle the coupling between the discrete IRS phase shift matrix and the BS beamforming matrix, we propose a series of transformations that convert the problem at hand to a more tractable mixed integer nonlinear programming (MINLP) problem.
    \item We develop an iterative algorithm based on the generalized Bender's decomposition (GBD) that obtains the globally optimal solution for the joint design problem and can serve as a performance benchmark for any corresponding suboptimal design, e.g., \cite{wu2019beamforming}, \cite{shi2022multiuser}. 
    \item Numerical results reveal that the proposed optimal joint design outperforms the commonly adopted suboptimal solution based on AO employing semidefinite relaxation. In particular, the performance gap between the proposed optimal design and the existing suboptimal design increases with the number of IRS elements.
\end{itemize}
\vspace*{-1.5mm}
The rest of this paper is organized as follows. In Section \rom{2}, we introduce the system model for the considered multiuser IRS-assisted wireless communication system with discrete phase shifters. In Section \rom{3}, the corresponding resource allocation problem is formulated. In Section \rom{4}, the optimal algorithm to solve the joint optimization problem for the BS beamforming vectors and the discrete IRS phase shifters is developed. Section \rom{5} evaluates the performance of the proposed optimal design via computer simulations, and Section \rom{6} concludes this paper.
\par
\textit{Notation:} 
Vectors and matrices are denoted by boldface lower case and boldface capital letters, respectively. $\mathbb{R}^{N\times M}$ and $\mathbb{C}^{N\times M}$ denote the space of $N\times M$ real-valued and complex-valued matrices, respectively. $||\cdot||_2$ denotes the $l_2$-norm of the argument. $(\cdot)^T$, $(\cdot)^*$, and $(\cdot)^H$ stand for the transpose, the conjugate, and the conjugate transpose of their arguments, respectively. $\mathbf{I}_{N}$ refers to the identity matrix of dimension $N$. $\mathrm{Tr}(\cdot)$ denotes the trace of the input argument. $\mathbf{0}_{1\times L}$ represents the $1\times L$ all-zeros row vector. $\mathbf{A}\succeq\mathbf{0}$ indicates that $\mathbf{A}$ is a positive semidefinite matrix. $\mathrm{diag}(\mathbf{a})$ denotes a diagonal matrix whose main diagonal elements are given by vector $\mathbf{a}$. $\mathrm{Re}\{\cdot\}$ and $\mathrm{Im}\{\cdot\}$ represent the real and imaginary parts of a complex number, respectively. $\mathbb{E}[\cdot]$ refers to statistical expectation. 
\section{IRS-Assisted Multiuser System Model}
\begin{figure}[t]
	\centering
	\includegraphics[width=3.0in]{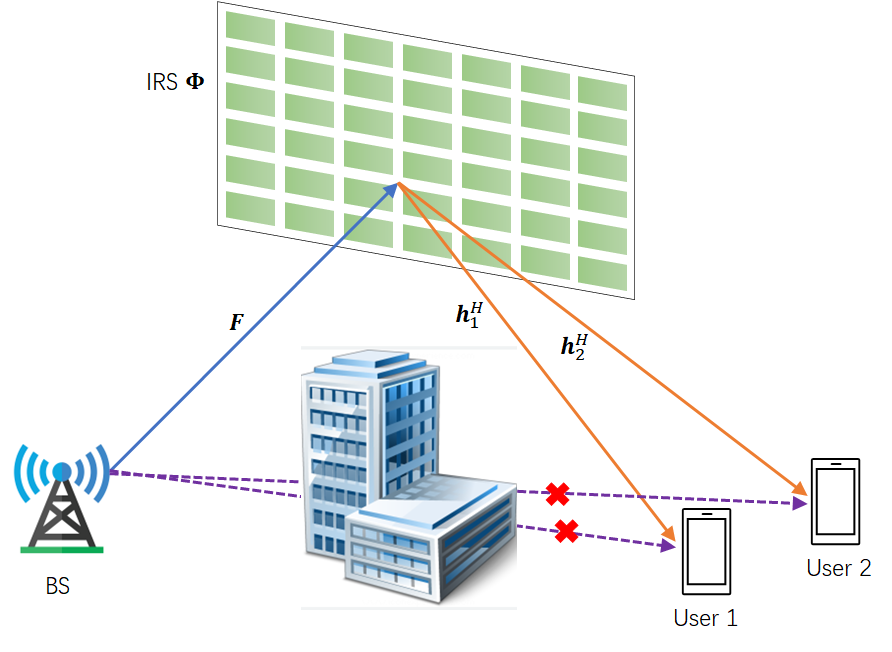}\vspace*{-5mm}
	\caption{An IRS-assisted multiuser MISO system comprising $K=2$ users.}\vspace*{-7mm}
	\label{fig::Model}
\end{figure}\vspace*{-2mm}

We consider a multiuser wireless communication system comprising a BS and $K$ users. The BS is equipped with $M$ antenna elements and serves $K$ single-antenna users. We study a scenario where the direct links from the BS to the $K$ users are blocked\footnote{At the expense of more involved notations, the considered framework can be extended to the case where direct communication links between the BS and the users exist.} by obstacles, e.g., buildings \cite{9014322}, \cite{huang2019reconfigurable}. To provide high-data-rate communication services for the $K$ users, an IRS comprising $N$ phase shift elements is deployed to establish strong reflecting paths between the BS and the $K$ users, see Fig. 1. In this paper, we consider the practical case where each element of the IRS admits only $L$ discrete phase shift values, i.e., values from the set $\{0,\Delta\theta,\cdots,(L-1)\Delta\theta\}$, where $\Delta\theta=\frac{2\pi}{L}$. Thus, the received baseband signal $y_k$ at the $k$-th user is given by
\vspace*{-2mm}
\begin{equation}\label{oriChannel}
    {y}_k=\mathbf{h}_k^H\bm{\Phi}\mathbf{F}\mathbf{W}\mathbf{s}+{n}_k,
    \vspace*{-2mm}
\end{equation}

where $\mathbf{h}_k\in\mathbb{C}^{N\times 1}$ denotes the conjugate channel from the IRS to the $k$-th user, and $\boldsymbol{\Phi}=\operatorname{diag}\left(e^{j\theta_1}.\cdots,e^{j\theta_N}\right)$ represents the phase shift matrix  of the IRS. The channel matrix from the BS to the IRS is denoted by $\mathbf{F}\in\mathbb{C}^{N\times M}$, and $\mathbf{W}=[\mathbf{w}_1,\cdots,\mathbf{w}_K]$ denotes the linear beamforming matrix at the BS, where $\mathbf{w}_k\in \mathbb{C}^{M\times 1}$ represents the linear beamforming vector for the $k$-th user. $\mathbf{s}=[s_1,\cdots,s_K]^T$ stands for the information-carrying symbol vector transmitted to the users, where $s_j\in\mathbb{C}$ denotes the symbol transmitted to the $j$-th user and $\mathbb{E}[|s_j|^2]=1,\mathbb{E}[s_j^*s_i]=0,j\neq i, \forall j,i\in\{1,\cdots,K\}$. $n_k\in\mathbb{C}$ represents the additive white Gaussian noise at the $k$-th user with zero mean and variance $\sigma_k^2$. For notational simplicity, we define sets $\mathcal{K}=\{1,\cdots,K\}$ and $\mathcal{N}=\{1,\cdots,N\}$ to collect the indices of the users and IRS elements, respectively.
\vspace*{-2mm}
\section{Optimization Problem Formulation}
\vspace*{-2mm}
To facilitate the optimal resource allocation design, we first define $\mathbf{X}=\bm{\Phi}\mathbf{F}\mathbf{W}, \mathbf{X}\in\mathbb{C}^{N\times K}$, and rewrite \eqref{oriChannel} as
\vspace*{-2mm}
\begin{equation}
    {y}_k=\mathbf{h}_k^H\sum_{j=1}^K\mathbf{x}_j{s}_j+{n}_k,\vspace*{-2mm}
\end{equation}
where $\mathbf{x}_j\in\mathbb{C}^{N\times 1}$ denotes the $j$-th column of $\mathbf{X}$. As a result, the signal-to-interference-plus-noise ratio (SINR) of user $k$ is given by
\vspace*{-2mm}
\begin{equation}
    \mathrm{SINR}_k=\frac{|\mathbf{h}_k^H\mathbf{x}_k|^2}{\sum_{k'\in\mathcal{K}\setminus\{k\}}|\mathbf{h}_k^H\mathbf{x}_{k'}|^2+\sigma_k^2}, \quad \forall k\in\mathcal{K}.\vspace*{-2mm}
\end{equation}
The design target of this paper is to minimize the total BS transmit power while guaranteeing a minimum required SINR for each user. The optimal BS beamforming policy and IRS phase shift configuration, i.e., $\mathbf{W}$ and $\bm{\Phi}$, are obtained by solving the following optimization problem
\vspace*{-2mm}
\begin{eqnarray}
\label{Ori_Problem1}
    &&\hspace*{-4mm}\underset{\mathbf{X},\mathbf{W},\bm{\Phi}}{\mino}\hspace*{2mm}\sum_{k\in\mathcal{K}}\left\|\mathbf{w}_k\right\|_2^2\notag\\
    &&\hspace*{2mm}\mbox{s.t.}\hspace*{7mm} \mbox{C1:}\hspace*{1mm} \frac{|\mathbf{h}_k^H\mathbf{x}_k|^2}{\sum_{k'\in\mathcal{K}\setminus\{k\}}|\mathbf{h}_k^H\mathbf{x}_{k'}|^2+\sigma_k^2}\geq \gamma_k,\hspace*{1mm}\forall k\in \mathcal{K},\notag\\
    &&\hspace*{14mm}\mbox{C2:}\hspace*{1mm}\mathbf{X}=\bm{\Phi}\mathbf{F}\mathbf{W},\notag\\
    &&\hspace*{14mm}\mbox{C3:}\hspace*{1mm}\theta_n\in\left\{0,\Delta\theta,\cdots,(L-1)\Delta\theta\right\}, \hspace*{1mm} \forall n\in \mathcal{N}.\vspace*{-5mm}
\end{eqnarray}
Here, $\gamma_k$ in constraint C1 denotes the minimum required SINR of the $k$-th user. Next, we define the phase shift vector $\bm{\theta}=[1,e^{j\Delta\theta },\cdots, e^{j(L-1)\Delta\theta}]^T$ and the binary selection vector of the $n$-th phase shifter $\mathbf{b}_n=\big[b_n[1],\cdots,b_n[L]\big]^T,\hspace*{1mm}\forall n$, where $b_n[l]\in\left\{0,\hspace*{1mm}1\right\},\hspace*{1mm}\forall l,n,\hspace*{2mm} \sum_{l=1}^{L}b_n[l]=1,\hspace*{1mm}\forall n$. Then, the coefficient of the $n$-th IRS phase shifter can be expressed as follows:
\vspace*{-1.5mm}
\begin{equation}
    e^{j\theta_n}=\mathbf{b}_n^T\bm{\theta}.\vspace*{-2mm}
\end{equation}
Hence, $\bm{\Phi}\in\mathbb{C}^{N\times N}$ can be written as
\vspace*{-2mm}
\begin{equation}
    \bm{\Phi}=\mathbf{B}\bm{\Theta},\vspace*{-2mm}
\end{equation}
where matrices $\bm{\Theta}\in\mathbb{C}^{NL\times N}$ and $\mathbf{B}\in\mathbb{C}^{N\times NL}$ are defined as follows, respectively,
\begin{eqnarray}
  \bm{\Theta}&\hspace*{-2mm}=\hspace*{-2mm}&
  \begin{bmatrix}
    \bm{\theta}^T & \mathbf{0}_{1\times L} & \mathbf{0}_{1\times L} & \cdots & \mathbf{0}_{1\times L}\\
    \mathbf{0}_{1\times L} & \bm{\theta}^T & \mathbf{0}_{1\times L} &\cdots & \mathbf{0}_{1\times L}\\
    \ldots & \ldots & \ldots & \ldots & \ldots\\
    \mathbf{0}_{1\times L} & \mathbf{0}_{1\times L} & \mathbf{0}_{1\times L} & \hspace*{1mm}\cdots & \bm{\theta}^T
  \end{bmatrix}^T,\\
  [2mm]
  \mathbf{B}&\hspace*{-2mm}=\hspace*{-2mm}&
  \begin{bmatrix}
    \mathbf{b}_1^T & \mathbf{0}_{1\times L} & \mathbf{0}_{1\times L} & \cdots & \mathbf{0}_{1\times L}\\
    \mathbf{0}_{1\times L} & \mathbf{b}_2^T & \mathbf{0}_{1\times L} &\cdots & \mathbf{0}_{1\times L}\\
    \ldots & \ldots & \ldots & \ldots & \ldots\\
    \mathbf{0}_{1\times L} & \mathbf{0}_{1\times L} & \mathbf{0}_{1\times L} & \hspace*{1mm}\cdots & \mathbf{b}_N^T
  \end{bmatrix}.
\end{eqnarray}
Then, we define $\widehat{\mathbf{H}}=\bm{\Theta}\mathbf{F},
\widehat{\mathbf{H}}\in\mathbb{C}^{NL\times M}$, and recast optimization problem \eqref{Ori_Problem1} equivalently as follows:
\vspace*{-2mm}
\begin{eqnarray}\label{Ori_Problem}
&&\hspace*{-4mm}\underset{\mathbf{X},\mathbf{W},\mathbf{B}}{\mino}\hspace*{2mm}\sum_{k\in\mathcal{K}}\left\|\mathbf{w}_k\right\|_2^2\notag\\
    &&\hspace*{2mm}\mbox{s.t.}\hspace*{7mm} \mbox{C1,}\notag\\
    &&\hspace*{14mm}\overline{\mbox{C2:}}\hspace*{1mm}\mathbf{X}=\mathbf{B}\widehat{\mathbf{H}}\mathbf{W},\notag\\
    &&\hspace*{14mm}\mbox{C3a:}\hspace*{1mm}\sum_{l=1}^{L}b_n[l]=1,\ \forall n,\notag\\
    &&\hspace*{14mm}\mbox{C3b:}\hspace*{1mm}b_n[l]\in\left\{0,\hspace*{1mm}1\right\},\hspace*{1mm}\forall l,\forall n.\vspace*{-3mm}
\end{eqnarray}
We note that due to the coupling between $\mathbf{W}$ and $\mathbf{B}$ and the binary feasible set introduced by constraint C3b, problem \eqref{Ori_Problem} is a combinatorial optimization problem which is NP-hard \cite{xu2022optimal}, \cite{wu2019beamforming}, \cite{shi2022multiuser}, \cite{hu2021robust}. Suboptimal algorithms for this non-convex combinatorial optimization problem have been developed based on the AO technique in \cite{wu2019beamforming}, \cite{shi2022multiuser}, but cannot reveal the best possible performance. In the next section, we develop a GBD-based iterative algorithm to optimally solve the resource allocation problem in \eqref{Ori_Problem} with guaranteed convergence.
\vspace*{-2mm}
\section{Optimal Design}
\vspace*{-2mm}
In this section, we aim to obtain the globally optimal solution to \eqref{Ori_Problem} by exploiting GBD theory. In the following, we first transform \eqref{Ori_Problem} into an equivalent MINLP problem, which paves the way for the development of the proposed GBD-based optimal algorithm.
\vspace*{-1.5mm}
\subsection{Problem Reformulation}
\vspace*{-2mm}
First, we reformulate the bilinear constraint $\overline{\mbox{C2}}$ to ensure that \eqref{Ori_Problem} satisfies the following two criteria which have to be met for the GBD framework to obtain the global optimal solution \cite{geoffrion1972generalized}:
\vspace*{-1mm}
\begin{itemize}
    \item Convexity: The given optimization problem is a convex optimization problem with respect to the continuous variables if the discrete variables are fixed.
    \item Linear separability: The given optimization problem is a standard linear programming problem with respect to the discrete variables if the continuous variables are fixed.\vspace*{-1mm}
\end{itemize}
As such, we transform equality constraint $\overline{\mbox{C2}}$ into two equivalent inequality constraints in the following lemma, whose proof is given in \cite[Appendix A]{6698281}.
\vspace*{-2mm}
\begin{lemma}
Equality constraint $\overline{\mbox{C2}}$ is equivalent to the following inequality constraints by introducing auxiliary optimization variables $\mathbf{S}\in\mathbb{C}^{N\times N}$ and $\mathbf{T}\in\mathbb{C}^{K\times K}$ and applying Schur's complement:
\vspace*{-2mm}
\begin{eqnarray}
\mbox{C2a:}&\hspace*{1mm}\label{sdp}
   \begin{bmatrix}
        \mathbf{S} & \mathbf{X} & \mathbf{B}\widehat{\mathbf{H}}\\
        \mathbf{X}^H & \mathbf{T} & \mathbf{W}^H\\
        \widehat{\mathbf{H}}^H\mathbf{B}^H & \mathbf{W} & \mathbf{I}_K
    \end{bmatrix}&\succeq \mathbf{0},\\
\mbox{C2b:}&\hspace*{1mm}\label{DC}
    \mathrm{Tr}\left(\mathbf{S}-\mathbf{B}\widehat{\mathbf{H}}\widehat{\mathbf{H}}^H\mathbf{B}\right)&\leq0,\vspace*{-2mm}
\end{eqnarray}
where $\mathbf{S}\succeq \mathbf{0}$ and $\mathbf{T}\succeq \mathbf{0}$.\vspace*{-2mm}
\end{lemma}
We note that constraint $\mbox{C2a}$ is a linear matrix inequality (LMI) constraint. 
Yet, constraint $\mbox{C2b}$ is in the form of a difference of convex functions, which is a non-convex constraint. To circumvent this obstacle, we define $\widebar{\mathbf{h}}_n=[\widehat{\mathbf{h}}_{(n-1)L+1}^T\widehat{\mathbf{h}}_{(n-1)L+1}^*,\cdots,\widehat{\mathbf{h}}_{nL}^T\widehat{\mathbf{h}}_{nL}^*]^T$, where $\widehat{\mathbf{h}}_{l}^T\in\mathbb{C}^{1\times M}$ denotes the $l$-th row of $\widehat{\mathbf{H}} $. Then, by exploiting the binary nature of matrix $\mathbf{B}$, constraint $\mbox{C2b}$ can be equivalently rewritten as the following affine constraint\vspace*{-2mm}
\begin{equation}\label{LPDC}
    \overline{\mbox{C2b}}\mbox{:}\hspace*{1mm}\mathrm{Tr}\big(\mathbf{S}\big)-\sum_{n=1}^N\widebar{\mathbf{h}}_n^T\mathbf{b}_n\leq 0.\vspace*{-2mm}
\end{equation}
On the other hand, SINR constraints C1 in \eqref{Ori_Problem} are also non-convex. Here, we observe that if any $\mathbf{x}_k, k\in\mathcal{K}$, satisfies the constraints in \eqref{Ori_Problem}, $e^{j\phi}\mathbf{x}_k$ with any arbitrary phase shift $\phi$ would also satisfy the constraints without altering the value of the objective function. Hence, we can transform the SINR constraints in C1 to convex constraints by using the following lemma.
\vspace*{-2mm}
\begin{lemma}
Without loss of generality and optimality, we assume that $\mathbf{h}_k^H\mathbf{x}_k\in\mathbb{R}$. Then, constraint C1 can be equivalently rewritten as
\vspace*{-2mm}
\begin{eqnarray}
&& \hspace*{-12mm}{{\mbox{C1a}}}\mbox{:}\hspace*{1mm}\sqrt{\sum_{k'\in\mathcal{K}\setminus\{k\}}|\mathbf{h}_k^H\mathbf{x}_{k'}|^2+\sigma_k^2}-\frac{\operatorname{Re}\{{\mathbf{h}}_{k}^H\mathbf{x}_k\}}{\sqrt{\gamma_k}}\leq 0, \hspace*{1mm}\forall k\in\mathcal{K},\\
&&\hspace*{-12mm}\mbox{C1b:}\hspace*{1mm}\operatorname{Im}\{{\mathbf{h}}_{k}^H\mathbf{x}_k\}=0,\hspace*{1mm}\forall k\in\mathcal{K}.\vspace*{-2mm}
\end{eqnarray}
C1a and C1b are both convex constraints.\vspace*{-2mm}
\end{lemma}
\vspace*{-2mm}
\begin{proof}
Please refer to \cite[Appendix \rom{2}]{zhang2008joint}.\vspace*{-2mm}
\end{proof}
Thus, the considered optimization problem can be equivalently transformed into
\vspace*{-2mm}
\begin{eqnarray}
\label{Reformulated_problem}
    &&\hspace*{-6mm}\underset{\substack{\mathbf{X},\mathbf{W},\mathbf{B},\\\mathbf{S},\mathbf{T}}}{\mino}\hspace*{2mm}\sum_{k\in\mathcal{K}}\left\|\mathbf{w}_k\right\|_2^2\notag\\
    &&\hspace*{0mm}\mbox{s.t.}\hspace*{8mm}\mbox{C1a},\mbox{C1b},\mbox{C2a},\overline{\mbox{C2b}},\mbox{C3a},\mbox{C3b}.\vspace*{-5mm}
\end{eqnarray}
Problem \eqref{Reformulated_problem} is an MILNP problem.
It can be verified that the optimization problem in \eqref{Reformulated_problem} satisfies the convexity and linear separability conditions that guarantee the global optimality of the GBD method. Considering this, next, we develop a GBD-based algorithm that has a succinct and generic structure\cite{ng2015secure}.
\vspace*{-6mm}
\subsection{GBD Procedure}
\vspace*{-2mm}
We develop a GBD-based iterative algorithm to optimally solve combinatorial optimization problem \eqref{Reformulated_problem}. In particular, we first decompose \eqref{Reformulated_problem} into a primal problem and a master problem. We obtain the primal problem by fixing the binary matrix $\mathbf{B}$ in \eqref{Reformulated_problem} and solve the primal problem with respect to (w.r.t.) $\mathbf{X}$, $\mathbf{W}$, $\mathbf{S}$, and $\mathbf{T}$, which produces an upper bound (UB) for the original problem in \eqref{Reformulated_problem}. Furthermore, by fixing the values of $\mathbf{X}$, $\mathbf{W}$, $\mathbf{S}$, and $\mathbf{T}$, the master problem is obtained in the form of a typical integer linear programming problem w.r.t. $\mathbf{B}$. The solution of the master problem is exploited to update the lower bound (LB) for the globally optimal objective function value of \eqref{Reformulated_problem}. By solving the primal problem and the master problem repeatedly, the GBD-based algorithm is guaranteed to converge to the globally optimal solution of \eqref{Reformulated_problem} as \eqref{Reformulated_problem} satisfies the convexity and linear separability conditions stated in Section \rom{3}-A \cite{geoffrion1972generalized}. In the following, we first formulate and solve the primal and master problems for the $i$-th iteration of the GBD algorithm. Then, we explain the overall GBD algorithm. 
\subsubsection{Primal Problem}
For given binary matrix $\mathbf{B}^{(i-1)}$, obtained by solving the master problem in the $(i-1)$-th iteration, the primal problem in the $i$-th iteration is given by

\vspace*{-6mm}
\begin{eqnarray}
\label{Primal_problem}
    &&\hspace*{-6mm}\underset{\mathbf{X},\mathbf{W},\mathbf{S},\mathbf{T}}{\mino}\hspace*{2mm}\sum_{k\in\mathcal{K}}\left\|\mathbf{w}_k\right\|_2^2\notag\\
    &&\hspace*{0mm}\mbox{s.t.}\hspace*{8mm} \mbox{C1a}, \mbox{C1b},\notag\\
    &&\hspace*{13mm}\mbox{C2a:}\hspace*{1mm}\begin{bmatrix}
    \mathbf{S} & \mathbf{X} & \mathbf{B}^{(i-1)}\widehat{\mathbf{H}} \\
    \mathbf{X}^H & \mathbf{T} & \mathbf{W}^H \\
    \widehat{\mathbf{H}}^H(\mathbf{B}^{(i-1)})^H & \mathbf{W} & \mathbf{I}_K \\
    \end{bmatrix}\succeq \mathbf{0},\notag\\
    &&\hspace*{13mm}\overline{\mbox{C2b}}:\hspace*{1mm} \mathrm{Tr}\big(\mathbf{S}\big)-\sum_{n=1}^N\widebar{\mathbf{h}}_n^T\mathbf{b}_n^{(i-1)}\leq 0.
   \vspace*{-6mm}
\end{eqnarray}
Note that problem \eqref{Primal_problem} is convex w.r.t. $\mathbf{X}$, $\mathbf{W}$, $\mathbf{S}$, and $\mathbf{T}$, and can be solved with standard convex program solvers such as CVX \cite{grant2008cvx}. The optimal solution of \eqref{Primal_problem} in the $i$-th iteration is denoted by $\mathbf{X}^{(i)},\mathbf{W}^{(i)},\mathbf{S}^{(i)}$, and $\mathbf{T}^{(i)}$. For the primal problem, we obtain the Lagrangian of \eqref{Primal_problem} as
\vspace*{-2mm}
\begin{equation}\label{Largrangian_conf}
    \begin{aligned}
            \mathcal{L}(\mathbf{X},\mathbf{W},\mathbf{S},\mathbf{T},\mathbf{B}^{(i-1)},\bm{\Lambda})&=\sum_{k\in\mathcal{K}}\left\|\mathbf{w}_k\right\|_2^2+f_1(\mathbf{X},\bm{\Lambda})\\
            &+f_2(\mathbf{X},\mathbf{W},\mathbf{S},\mathbf{T},\mathbf{B}^{(i-1)},\bm{\Lambda}),
    \end{aligned}\vspace*{-2mm}
\end{equation}
where 
\vspace*{-2mm}
\begin{equation}\notag
\begin{aligned}
    f_1(\mathbf{X},\bm{\Lambda})&=\sum_{k\in\mathcal{K}}\alpha_k\left(\sqrt{\sum_{k'\in\mathcal{K}\setminus\{k\}}\hspace*{-2mm}|\mathbf{h}_k^H\mathbf{x}_{k'}|^2+\sigma_k^2}-\frac{\operatorname{Re}\{\mathbf{h}_k^H\mathbf{x}_k\}}{\sqrt{\gamma_k}}\right)\\
    &+\sum_{k\in\mathcal{K}}\beta_k\big(\operatorname{Im}\{\mathbf{h}_k^H\mathbf{x}_k\}\big),
\end{aligned}\vspace*{-2mm}
\end{equation}
\vspace*{-3mm}
\begin{eqnarray}\notag
     f_2(\mathbf{X},\mathbf{W},\mathbf{S},\mathbf{T},\mathbf{B}^{(i-1)},\bm{\Lambda})&\hspace*{-2.5mm}=\hspace*{-2.5mm}&\mathrm{Tr}\big(\mathbf{S}\mathbf{Q}_{11}\big)+\mathrm{Tr}\big(\mathbf{T}\mathbf{Q}_{22}\big)\notag\\
&\hspace*{-2.5mm}+\hspace*{-2.5mm}&q\Big(\mathrm{Tr}\big(\mathbf{S}\big)-\sum_{n=1}^N\widebar{\mathbf{h}}_n^T\mathbf{b}^{(i-1)}_n\Big)\notag\\
    &\hspace*{-2.5mm}+\hspace*{-2.5mm}&2\mathrm{Re}\left\{\mathrm{Tr}\big({\mathbf{Q}^H_{32}}\mathbf{W}\big)+\mathrm{Tr}\big(\mathbf{X}\mathbf{Q}_{21}\big)\right\}\notag\\
    &\hspace*{-2.5mm}+\hspace*{-2.5mm}&2\mathrm{Re}\left\{\mathrm{Tr}\big(\mathbf{B}^{(i-1)}\hat{\mathbf{H}}\mathbf{Q}_{31}\big)\right\}\notag.\vspace*{-2mm}
\end{eqnarray}
Here, $\bm{\Lambda}=\left\{\alpha_k,\beta_k,\mathbf{Q},{q}\right\}$ denotes the collection of dual variables, where $\alpha_k,\beta_k,\mathbf{Q}$, and ${q}$ represent the dual variables for constraints C1a, C1b, C2a, and C2b, respectively. The dual variable matrix $\mathbf{Q}\in\mathbb{C}^{(N+K+M)\times (N+K+M)}$ for constraint C2a is decomposed as
\vspace*{-2mm}
\begin{equation}\label{Qi}
  \mathbf{Q}=\left[ \begin{array}{ccc}
        \mathbf{Q}_{11} & \mathbf{Q}_{21}^H & \mathbf{Q}_{31}^H\\
        \mathbf{Q}_{21} & \mathbf{Q}_{22} & \mathbf{Q}_{32}^H\\
        \mathbf{Q}_{31} & \mathbf{Q}_{32} & \mathbf{Q}_{33}
    \end{array}\right],\vspace*{-2mm}
\end{equation}
where $\mathbf{Q}_{11}\in\mathbb{C}^{N\times N}$, $\mathbf{Q}_{21}\in\mathbb{C}^{K\times N}$, $\mathbf{Q}_{22}\in\mathbb{C}^{K\times K}$, $\mathbf{Q}_{31}\in\mathbb{C}^{M\times N}$, $\mathbf{Q}_{32}\in\mathbb{C}^{M\times K}$, and $\mathbf{Q}_{33}\in\mathbb{C}^{M\times M}$. We define $\bm{\Lambda}^{(i)}$ 
 as the the optimal dual solution of \eqref{Primal_problem} in the $i$-th iteration. If problem \eqref{Primal_problem} is infeasible for a given $\mathbf{B}^{(i-1)}$, then we formulate an $l_1$-minimization feasibility-check problem as follows
 \vspace*{-2mm}
\begin{eqnarray}
\label{Feasible_problem}
    &&\hspace*{-6mm}\underset{\mathbf{X},\mathbf{W},\mathbf{S},\mathbf{T},\bm{\lambda}}{\mino}\hspace*{2mm}\sum_{k\in\mathcal{K}}\lambda_k\notag\\
    &&\hspace*{0mm}\mbox{s.t.}\hspace*{8mm}\mbox{C1b},\mbox{C2a},\overline{\mbox{C2b}}, \notag\\
    &&\hspace*{13mm}\mbox{C1a}\mbox{:}\hspace*{1mm}\sqrt{\sum_{k'\neq k}|\mathbf{h}_k^H\mathbf{x}_{k'}|^2+\sigma_k^2}-\frac{\operatorname{Re}\{{\mathbf{h}}_{k}^H\mathbf{x}_k\}}{\sqrt{\gamma_k}}\leq \lambda_k,\notag\\
    &&\hspace*{21mm}\forall k\in \mathcal{K},\notag\\
    &&\hspace*{13mm}\mbox{C4}\mbox{:}\hspace*{1mm}\lambda_k\geq 0, \forall k\in\mathcal{K},\vspace*{-2mm}
\end{eqnarray}
where $\bm{\lambda}=[\lambda_1,\cdots,\lambda_K]$ is an auxiliary optimization variable. Problem \eqref{Feasible_problem} is convex, always feasible, and can be solved with CVX \cite{grant2008cvx}. Similar to the notation in \eqref{Primal_problem}, we define $\widetilde{\bm{\Lambda}}=[\widetilde{\alpha}_k,\widetilde{\beta_k},\widetilde{\mathbf{Q}},\widetilde{q}]$, where $\widetilde{\alpha}_k,\widetilde{\beta}_k,\widetilde{\mathbf{Q}}$, and $\widetilde{q}$ are the dual variables for constraints C1a, C1b, C2a, and $\overline{\mbox{C2b}}$ in \eqref{Feasible_problem}, respectively, and the optimal solutions of \eqref{Feasible_problem} are denoted by $\widetilde{\mathbf{X}}^{(i)},\widetilde{\mathbf{W}}^{(i)},\widetilde{\mathbf{S}}^{(i)}$, and $\widetilde{\mathbf{T}}^{(i)}$. The Lagrangian of \eqref{Feasible_problem} is given by
\vspace*{-2mm}
\begin{equation}\label{Largrangian_feasible}
\begin{aligned}
    \widetilde{\mathcal{L}}(\mathbf{X},\mathbf{W},\mathbf{S},\mathbf{T},\mathbf{B}^{(i-1)},\widetilde{\bm{\Lambda}})&={f}_1(\mathbf{X},\widetilde{\bm{\Lambda}})\\
    &+{f}_2(\mathbf{X},\mathbf{W},\mathbf{S},\mathbf{T},\mathbf{B}^{(i-1)},\widetilde{\bm{\Lambda}}),
\end{aligned}\vspace*{-2mm}
\end{equation}
 We define $\widetilde{\bm{\Lambda}}^{(i)}$ as the the optimal dual solution of \eqref{Primal_problem} in the $i$-th iteration. The solutions of the feasibility-check problem \eqref{Feasible_problem} are used to generate a feasibility cut, i.e.,\\ $0\geq\min_{\widetilde{\mathbf{X}},\widetilde{\mathbf{W}},\widetilde{\mathbf{S}},\widetilde{\mathbf{T}}}\widetilde{\mathcal{L}}(\widetilde{\mathbf{X}},\widetilde{\mathbf{W}},\widetilde{\mathbf{S}},\widetilde{\mathbf{T}},\mathbf{B},\widetilde{\bm{\Lambda}}^{(t)})$, separating the infeasible solution $\mathbf{B}^{(i-1)}$ from the feasible set of the master problem in the following iterations.
\subsubsection{Master Problem}
The master problem is derived based on nonlinear convex duality theory \cite{geoffrion1972generalized}. Let $\mathcal{F}^{(i)}$ and $\mathcal{I}^{(i)}$ denote the sets collecting the iteration indices for which \eqref{Primal_problem} is feasible and infeasible before solving the master problem in the $i$-th iteration, respectively. Then, we recast the master problem in the $i$-th iteration in its epigraph form by introducing auxiliary optimization variable $\eta$, resulting in
\vspace*{-2mm}
\begin{eqnarray}
\label{Master_problem}
    &&\hspace*{-8mm}\underset{\mathbf{B},\eta}{\mino}\hspace*{2mm}\eta\notag\\
    &&\hspace*{-4mm}\mbox{s.t.}\hspace*{2mm}\mbox{C3a}, \mbox{C3b},\notag\\
    &&\hspace*{2mm}\mbox{C5a}\mbox{:}\hspace*{1mm}\eta\geq\min_{\substack{\mathbf{X},\mathbf{W},\\\mathbf{S},\mathbf{T}}}\mathcal{L}(\mathbf{X},\mathbf{W},\mathbf{S},\mathbf{T},\mathbf{B},\bm{\Lambda}^{(t)}),\hspace*{0mm}\forall t\in\mathcal{F}^{(i)},\notag\\
    &&\hspace*{2mm}\mbox{C5b}\mbox{:}\hspace*{1mm}0\geq\min_{\substack{\widetilde{\mathbf{X}},\widetilde{\mathbf{W}},\\\widetilde{\mathbf{S}},\widetilde{\mathbf{T}}}}\widetilde{\mathcal{L}}(\widetilde{\mathbf{X}},\widetilde{\mathbf{W}},\widetilde{\mathbf{S}},\widetilde{\mathbf{T}},\mathbf{B},\widetilde{\bm{\Lambda}}^{(t)}),\hspace*{0mm}\forall t\in\mathcal{I}^{(i)},\vspace*{-2mm}
\end{eqnarray}
where constraints C5a and C5b are referred to as the optimality and feasibility cuts \cite{geoffrion1972generalized}, respectively. It is not hard to verify that the global minimum over $\mathbf{X},\mathbf{W},\mathbf{S},\mathbf{T}$ of $\mathcal{L}(\mathbf{X},\mathbf{W},\mathbf{S},\mathbf{T},\mathbf{B},\bm{\Lambda})$ can be obtained independently of $\mathbf{B}$ for fixed $\bm{\Lambda}$. Thus, the solutions of $\underset{\mathbf{X},\mathbf{W},\mathbf{S},\mathbf{T}}{\min}\hspace*{2mm}\mathcal{L}(\mathbf{X},\mathbf{W},\mathbf{S},\mathbf{T},\mathbf{B},\bm{\Lambda}^{(i)})$ in \eqref{Master_problem} are the solutions of \eqref{Primal_problem} obtained in the $i$-th iteration based on duality theory. Similarly, we can verify that the solutions of  $\underset{\widetilde{\mathbf{X}},\widetilde{\mathbf{T}},\widetilde{\mathbf{S}},\widetilde{\mathbf{T}}}{\min}\hspace*{2mm}\widetilde{\mathcal{L}}(\widetilde{\mathbf{X}},\widetilde{\mathbf{W}},\widetilde{\mathbf{S}},\widetilde{\mathbf{T}},\mathbf{B},\widetilde{\bm{\Lambda}}^{(i)})$ in \eqref{Master_problem} are the solutions of \eqref{Feasible_problem} if the primal problem is not feasible in the $i$-th iteration. By substituting $\mathbf{X}^{(t)},\mathbf{W}^{(t)},\mathbf{S}^{(t)},\mathbf{T}^{(t)}, \forall t \in \mathcal{F}^{(i)}$ and $\widetilde{\mathbf{X}}^{(t)},\widetilde{\mathbf{W}}^{(t)},\widetilde{\mathbf{S}}^{(t)}$, $\widetilde{\mathbf{T}}^{(t)}, \forall t \in \mathcal{I}^{(i)}$ in C5a and C5b, respectively, the master problem in \eqref{Master_problem} becomes a mixed integer linear programming (MILP) problem, which can be optimally solved by employing standard numerical solvers for MILPs, e.g.,  MOSEK \cite{grant2008cvx}. The master problem provides a lower bound, $\eta^{(i)}$, to original problem \eqref{Reformulated_problem} and its solution, $\mathbf{B}^{(i)}$, is used to generate the primal problem in the next iteration. 
\setlength{\textfloatsep}{0pt}
\begin{algorithm}[t]
\caption{Optimal Resource Allocation Algorithm}
\begin{algorithmic}[1]
\small
\STATE Set iteration index $i=0$, initialize upper bound $\mathrm{UB}^{(0)}\gg 1$, lower bound $\mathrm{LB}^{(0)}=0$, the set of the feasible iterations indices $\mathcal{F}^{(0)}=\emptyset$, the set of the infeasible iterations indices $\mathcal{I}^{(0)}=\emptyset$, and convergence tolerance $\Delta\ll 1$, generate a feasible $\mathbf{B}^{(0)}$
\REPEAT
\STATE Set $i=i+1$
\STATE Solve \eqref{Primal_problem} for given $\mathbf{B}^{(i-1)}$.
\IF{the primal problem \eqref{Primal_problem} is feasible}
\STATE Update $\mathbf{X}^{(i)},\mathbf{W}^{(i)},\mathbf{S}^{(i)}$, and $\mathbf{T}^{(i)}$ and store the corresponding objective function value of $\sum_{k\in\mathcal{K}}\left\|\mathbf{w}_k^{(i)}\right\|_2^2$
\STATE Construct $\mathcal{L}(\mathbf{X},\mathbf{W},\mathbf{S},\mathbf{T},\mathbf{B},\bm{\Lambda}^{(i)})$ based on \eqref{Largrangian_conf}
\STATE Update the upper bound of \eqref{Reformulated_problem} based on $\mathrm{UB}^{(i)}=\min\left\{\mathrm{UB}^{(i-1)},\hspace*{1mm}\sum_{k\in\mathcal{K}}\left\|\mathbf{w}_k^{(i)}\right\|_2^2\right\}$ 
\STATE Update $\mathcal{F}^{(i)}$ by $\mathcal{F}^{(i-1)}\cup\{i\}$ and $\mathcal{I}^{(i)}=\mathcal{I}^{(i-1)}$
\ELSE
\STATE Solve \eqref{Feasible_problem} for given $\mathbf{B}^{(i-1)}$, update $\widetilde{\mathbf{X}}^{(i)}$, $\widetilde{\mathbf{W}}^{(i)}$, $\widetilde{\mathbf{S}}^{(i)}$, $\widetilde{\mathbf{T}}^{(i)}$
\STATE Construct $\widetilde{\mathcal{L}}(\mathbf{X},\mathbf{W},\mathbf{S},\mathbf{T},\mathbf{B},\widetilde{\bm{\Lambda}}^{i})$ based on \eqref{Largrangian_feasible}
\STATE Update $\mathcal{I}^{(i)}$ by $\mathcal{I}^{(i-1)}\cup\{i\}$ and $\mathcal{F}^{(i)}=\mathcal{F}^{(i-1)}$
\ENDIF
\STATE Solve the relaxed master problem \eqref{Master_problem} and update $\eta^{(i)}$ and $\mathbf{B}^{(i)}$
\STATE Update the lower bound as $\mathrm{LB}^{(i)}=\eta^{(i)}$
\UNTIL $\mathrm{UB}^{(i)}-\mathrm{LB}^{(i)}\leq \Delta$
\end{algorithmic}
\end{algorithm}

\subsubsection{Overall Algorithm}
The entire GBD procedure is summarized in \textbf{Algorithm 1}. Index $i$ is first set to zero and binary matrix $\mathbf{B}$ is initialized. In the $i$-th iteration, we first solve problem \eqref{Primal_problem}. If \eqref{Primal_problem} is feasible, we obtain the intermediate solutions for $\mathbf{X}^{(i)}$, $\mathbf{W}^{(i)}$, $\mathbf{S}^{(i)}$, $\mathbf{T}^{(i)}$ and the corresponding Lagrangian multiplier set $\bm{\Lambda}^{(i)}$ to generate the optimality cut. Besides, the objective value $\sum_{k\in\mathcal{K}}\left\|\mathbf{w}_k^{(i)}\right\|_2^2$ obtained in the $i$-th iteration is used to update the performance upper bound $\mathrm{UB}^{(i)}$. If \eqref{Primal_problem} is infeasible, we turn to solve the feasibility-check problem in \eqref{Feasible_problem}. The intermediate solutions for $\widetilde{\mathbf{X}}^{(i)}$, $\widetilde{\mathbf{W}}^{(i)}$, $\widetilde{\mathbf{S}}^{(i)}$, $\widetilde{\mathbf{T}}^{(i)}$ and the corresponding Lagrangian multiplier set $\widetilde{\bm{\Lambda}}^{(i)}$ are used to generate the infeasibility cut. Then, the master problem in \eqref{Master_problem} is optimally solved using a standard MILP solver. The objective value of the master problem provides a performance lower bound $\mathrm{LB}^{(i)}$ for the original optimization problem in \eqref{Ori_Problem}. Following the above procedure, we can gradually reduce the gap between $\mathrm{LB}$ and $\mathrm{UB}$
in each iteration. It has been shown in \cite[Theorem 2.4]{geoffrion1972generalized} that GBD-based algorithms are guaranteed to converge to the globally optimal solution of MINLP problems in a finite number of iterations for a given convergence tolerance $\Delta\geq 0$, if the convexity and linear separability conditions are satisfied, as is the case for the problem at hand. Although the worst case computational complexity of the proposed GBD-based algorithm scales exponentially with the number of IRS elements, in our simulation experiments, the proposed GBD method requires significantly fewer iterations to converge compared with an exhaustive search.
\vspace*{-2mm}
\section{Numerical Results}
\vspace*{-2mm}
In this section, we evaluate the performance of the proposed optimal scheme via simulations. Specifically, we consider a system with a BS equipped with $M=6$ transmit antennas to provide communication services to $K=4$ users. As the direct links between the BS and the users are blocked, an IRS is deployed. In particular, the IRS is $D=25$ m away from the BS and the users are evenly distributed on a half-circle with the center being the location of the IRS and radius $r=10$ m. The noise variances of all users are set to $\sigma_k^2=-117$ dBm, $\forall k \in\mathcal{K}$ \cite{guo2020weighted}. All channels in the considered network are assumed to be Rician distributed. The channel matrix $\mathbf{F}$ between BS and IRS is modeled as follows
\begin{equation}\vspace*{-1mm}
    \mathbf{F}=\sqrt{L_0D^{-\alpha_{\mathrm{BI}}}}\left(\sqrt{\frac{\beta_{\mathrm{BI}}}{1+\beta_{\mathrm{BI}}}}\mathbf{F}_{\mathrm{L}}+\sqrt{\frac{1}{1+\beta_{\mathrm{BI}}}}\mathbf{F}_{\mathrm{N}}\right),
\end{equation}\vspace*{-1mm}
where $L_0$ denotes the large-scale fading at reference distance $d_0=1$ m. Parameters $\alpha_{\mathrm{BI}}=2.2$ and $\beta_{\mathrm{BI}}=1$ denote the path loss exponent and the Rician factor, respectively. Matrices $\mathbf{F}_{\mathrm{L}}$ and $\mathbf{F}_{\mathrm{N}}$ represent the deterministic line-of-sight (LoS) and random non-LoS (NLoS) components, respectively. $\mathbf{F}_{\mathrm{L}}$ is generated by computing the product of the receive and transmit array response vectors while the NLoS component $\mathbf{F}_{\mathrm{N}}$ is assumed to follow a Rayleigh distribution. The channel vector for the IRS-user $k$ link, i.e., $\mathbf{h}_k$, is generated in a similar manner as $\mathbf{F}$. The path loss exponent and the Rician factor of $\mathbf{h}_k$, $\forall k \in\mathcal{K}$, i.e., $\alpha_{\mathrm{IU}}^k$ and $\beta_{\mathrm{IU}}^k$, are set to $2.8$ and $1$, respectively. For ease of presentation, we assume all users impose the same minimum SINR requirement, i.e., $\gamma_k=\gamma, \forall k\in\mathcal{K}$. We consider a 2-bit phase shifter for all reflecting elements of the IRS, i.e., $L=4$. The convergence tolerance of \textbf{Algorithm 1} is set as $\Delta=0$ to obtain the globally optimal solution. The number of channel realizations considered in the following simulations is $100$.
\begin{figure}[t]\vspace*{-5mm}
	\centering
	\includegraphics[width=3.2in]{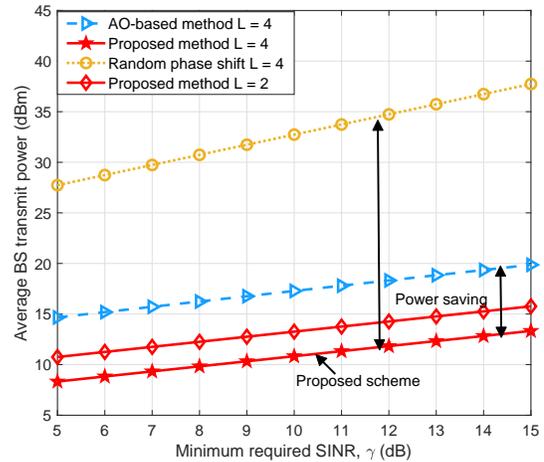}\vspace*{-5mm}
	\caption{Average BS transmit power versus the minimum required SINR of the users for different IRS quantization resolutions.}
	\label{fig::SINR}
\end{figure}
For comparison, we consider two baseline schemes. For baseline scheme 1, the IRS phase shifts are randomly selected from the set of discrete phase shift values and the beamforming vector $\mathbf{w}_k$ is designed by solving the classic beamforming problem in wireless systems using the semidefinite relaxation method. Furthermore, we adopt the AO-based algorithm as the second baseline scheme \cite{9014322}, \cite{wu2019intelligent}, which optimizes the BS beamforming matrix and continuous IRS phase shifts alternatingly in an iterative manner by adopting the semidefinite relaxation technique until convergence. After convergence, the obtained continuous phase shifts are quantized to the feasible discrete phase shift values.

In Fig. \ref{fig::SINR}, we show the average BS transmit power versus the minimum required SINR values of the users for $N=64$ phase shifters. For the considered SINR range and $L=4$, the proposed scheme requires on average approximately $180$ iterations to find the global optimal solution, which is much faster than an exhaustive search over all $4^{64}$ possible IRS phase shift configurations. As can be observed, for the proposed scheme and the two baseline schemes, the required transmit power at the BS increases monotonically with the minimum required SINR value. This is attributed to the fact that to meet more rigorous QoS requirements of the users, the BS has to increase the transmit power for beamforming. Moreover, we observe that the proposed scheme outperforms the two baseline schemes for the entire considered range of $\gamma$. In particular, baseline scheme 1 employs a randomly generated phase shift pattern, which cannot focus the reflected signal power towards the users. As for baseline scheme 2, the phase shifts of the IRS and the beamforming matrix obtained with the AO algorithm may get stuck in a stationary point close to the initial point resulting in a limited performance improvement over the iterations\cite{xu2021resource}. This reveals the importance of optimizing all the available DoFs in IRS-assisted wireless systems jointly and optimally. Moreover, we observe that for the proposed optimal scheme, replacing the 2-bit phase shifters, i.e., $L=4$, with cheaper 1-bit on-off phase shifters, i.e., $L=2$, leads to roughly a $2.5$ dB power loss, which suggests a trade-off between the transmit power and phase shifter cost in IRS-assisted communication systems.
\begin{figure}[t]\vspace*{-5mm}
	\centering
	\includegraphics[width=3.2in]{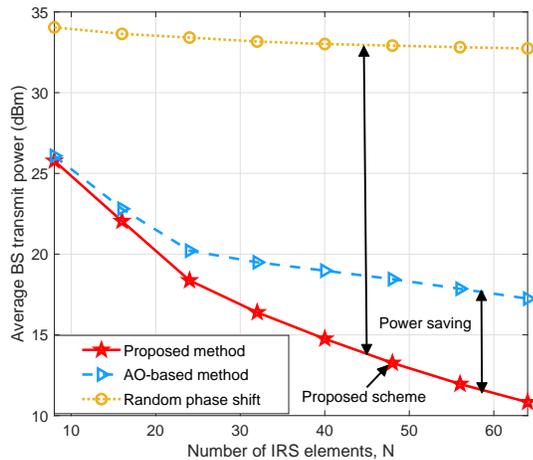}\vspace*{-5mm}
	\caption{Average BS transmit power versus the number of IRS elements with $L=4$.}
	\label{fig::N}
\end{figure}
Fig. \ref{fig::N} shows the BS transmit power versus the number of IRS elements for $\gamma=10$ dB. As can be seen from the figure, the BS transmit powers of the proposed scheme and the two baseline schemes decrease as $N$ increases. This can be explained as follows. First, a larger number of IRS elements can reflect more power towards the desired users leading to a potential power gain. Second, the extra phase shifters contribute additional DoFs for system design, which can be exploited to customize a more favourable wireless channel for effective information transmission. Also, we observe that compared with the proposed optimal scheme, both baseline schemes require a higher transmit power. Specifically, baseline scheme 1 adopting random phase shifts can only achieve small IRS array gain but fails to fully utilize the DoFs offered by the IRS for energy-focused beamforming. As for baseline scheme 2, we observe that for a small-scale IRS, e.g., $N=8$, there is only a relatively small performance loss compared to the proposed scheme. In other words, the AO-based algorithm employed in baseline scheme 2 may converge to a high-quality solution of the considered optimization problem for a small-scale IRS since the corresponding feasible set is relatively small. Yet, as $N$ increases to a typical value \cite{9014322}, e.g., $N\geq 40$, the gap between baseline scheme 2 and the proposed scheme is enlarged. In fact, the feasible solution set grows rapidly with $N$ such that the AO-based algorithm is more likely to get stuck in an unsatisfactory stationary point for larger values of $N$, leading to a significantly higher BS transmit power (roughly $7.5$ dB additional power consumption for $N=64$) compared with the proposed scheme that always finds the globally optimal solution. 
\vspace*{-2.5mm}
\section{Conclusion}
\vspace*{-2mm}
In this paper, we studied the optimal resource allocation algorithm design for multiuser IRS-assisted systems with discrete phase shifts at the IRS. We formulated the resource allocation design as a non-convex MILNP problem for the minimization of the BS transmit power subject to given minimum SINR requirements of the users. An optimal iterative GBD-based algorithm was developed to obtain the global optimal solution of the proposed resource allocation problem. The simulation results revealed that the proposed algorithm can significantly reduce the required transmit power compared with the state-of-the-art AO-based suboptimal algorithm. In future work, we will consider a more general scenario, where a direct link between the BS and the users exists and the channel state information is not perfectly known, and develop corresponding computationally efficient algorithms.
\vspace*{-3mm}
    
\bibliographystyle{IEEEtran}
\bibliography{Reference_List}
\end{document}